# Motor control by precisely timed spike patterns

Kyle H. Srivastava[1,*], Caroline M. Holmes[2,*], Michiel Vellema[3], Andrea Pack[4], Coen P. H. Elemans[3], Ilya Nemenman[2], and Samuel J. Sober[5,6]

[1]Biomedical Engineering Doctoral Program, Georgia Institute of Technology and Emory University, Atlanta, GA, USA

[2]Departments of Physics and Biology, Emory University, Atlanta, GA, USA

[3]Department of Biology, University of Southern Denmark, Odense, Denmark

[4]Neuroscience Doctoral Program and Department of Biology, Emory University, Atlanta, GA, USA

[5]Department of Biology, Emory University, Atlanta, GA, USA

[6]Corresponding author (email: samuel.j.sober@emory.edu)

[*]Denotes equal contributions

**Summary:**

A fundamental problem in neuroscience is to understand how sequences of action potentials ("spikes") encode information about sensory signals and motor outputs. Although traditional theories of neural coding assume that information is conveyed by the total number of spikes fired (spike rate), recent studies of sensory [1-5] and motor [6] activity have shown that far more information is carried by the millisecond-scale timing patterns of action potentials (spike timing). However, it is unknown whether or how subtle differences in spike timing drive differences in perception or behavior, leaving it unclear whether the information carried by spike timing actually plays a causal role in brain function [1]. Here we demonstrate how a precise spike timing code is read out downstream by the muscles to control behavior. We provide both correlative and causal evidence to show that the nervous system uses millisecond-scale variations in the timing of spikes within multi-spike patterns to regulate a relatively simple behavior – respiration in the Bengalese finch, a songbird. These findings suggest that a fundamental assumption of current theories of motor coding requires revision, and that significant improvements in applications, such as neural prosthetic devices, can be achieved by using precise spike timing information.

**Main Text**

The brain uses sequences of spikes to encode sensory and motor signals. In principle, neurons can encode this information via their firing rates, the precise timing of their spikes, or both [3, 5]. While many studies have shown that spike timing contains information beyond that in the rate in sensory codes [1, 2, 4], these studies could not verify whether precise timing affects perception or behavior. In motor systems, rate coding approaches dominate [7, 8], but we recently showed that precise cortical spike timing contains much more information about upcoming behavior than does rate [6]. However, as in sensory systems, it remains unclear whether spike timing actually *controls* variations in behavior.

A spike timing-based theory of motor production predicts that millisecond-scale fluctuations in spike timing, holding other spike train features constant, will causally influence behavior. We tested this prediction by focusing largely on the minimal patterns that have variable spike timing but fixed firing rate, burst onset, and burst duration: sequences of three spikes ("triplets"), where the third spike is a fixed latency after the first, but the timing of the middle spike varies. We examined timing codes in songbirds by focusing on respiration, which offers two key advantages. First, breathing is a relatively slow behavior (cycles last ~400-1000 msec), so the existence of timing codes is not *a priori* necessary, and yet the precise control of breathing during singing [9, 10] suggests that timing *may* play a role. Second, a novel electrode system allowed us to collect spiking data over >50,000 breaths, yielding the large data set sizes necessary to decipher the neural code.

We recorded electromyographic (EMG) signals from the expiratory muscle group (EXP) [11] using a flexible microelectrode array (Figure 1a) to isolate spikes from single motor units (that is, the muscle fibers innervated by a single motor neuron). Precise timing codes might be implemented by individual spikes [3, 12, 13] or by timing of spikes within a multispike pattern. Thus we first verified whether single motor unit spike trains contain multispike features at high temporal resolution. Analysis of inter-spike intervals (ISIs) revealed that spiking was more regular than expected from a Poisson process (see Extended Data), demonstrating that the nervous system precisely regulates spike timing and suggesting that timing can be used to control behavior.

Next we quantified the timescale on which the nervous system controls spikes within triplets by measuring the mutual information between consecutive ISIs in anesthetized birds. The nonzero value of information (Figure 1c, red dot) suggests that consecutive ISIs (and hence spike triplets) are controlled in the neural code. To understand the characteristic timescale of this control, we jittered the timing of each ISI by a Gaussian random number with standard deviation $\sigma$ and again estimated the consecutive ISIs mutual information. We found that the information only approaches its un-jittered values for $\sigma \sim 1$ ms (Figure 1c, blue), demonstrating that spike trains have millisecond-scale features. Similar findings were obtained in awake birds (Extended Data Figure 1).

We then asked whether these millisecond-scale features predict behavior by simultaneously recording single motor units and air pressure within the respiratory system. Since respiration is controlled by ensembles of motor units, we expected a single motor neuron not to drive the breathing cycle, but only fluctuations around the mean. We therefore subtracted the mean respiratory pressure waveform from the recorded pressure (Figure 1a) and investigated the relationship between such pressure residuals and the preceding spike train (Figure 2a) using a novel estimator of mutual information (see Methods) [14]. This method separates the total mutual information between spikes and pressure residuals into contributions from spike count and spike timing:

$$I(spikes, pressure) = I(spike\ count, pressure) + I(spike\ timing, pressure).$$

Seven of eight birds tested (all but EMG3) had statistically significant information in spike timing, which was of the same order as the information in spike rate (Figure 2a), indicating that precise spike timing in motor units predicts the ensuing behavior.

We then verified directly that specific spike patterns predict behavior by selecting all pressure residuals that followed particular patterns and evaluating their means (pattern-triggered averages, or PTAs) and variances. Specifically, we focused on triplets preceded by $\geq 30$ ms of silence, where the first and third spikes occurred 20 ms apart (at 2 ms accuracy), and which differed only by the timing of the middle spike (10 ms vs. 12 ms after the first spike; or "10-10" and "12-8" triplets, respectively; green and blue marks in Figure 2b). Such patterns had identical firing rates (3 spikes in 20 ms) and burst onset/offset times, and were among the most common observed patterns (e.g., $N = 23{,}991$ and $11{,}558$ for the two patterns in Bird EMG1, or 11% and 5% of all spike triplets of $\leq 20$ ms duration).

We found that PTAs following the 10-10 and 12-8 spike triplets were significantly different (Figure 2b). We quantified discriminability of the PTAs using the *d'* statistic [15] and found $d' = 0.108 \pm 0.011$ (s.d.) 17 ms after triplet onset (Figure 2c). The same effect is present across all six birds (Figure 2d). Notably, although *d'* traces are similar across animals, the PTAs themselves are not (Extended Data Figure 2). Therefore, while the discriminability of different patterns is consistent (Figure 2d), the encoding of pressure differs across individuals. Wavelet-based functional ANOVA (wfANOVA, see *Methods*) revealed a consistent significant effect between the PTAs across birds after accounting for intersubject variability (Extended Data Figure 3, Extended Data Table 1). Therefore, millisecond-scale changes in timing of a single spike in a multispike pattern at a fixed firing rate predict significant changes in air sac pressure. This agrees with our previous findings that cortical neurons upstream of vocal and respiratory muscles also use spike timing to encode behavior [6].

Although the above results demonstrate that precise spike timing predicts behavioral variations, they cannot reveal whether timing *affects* muscle output. To test this, we extracted muscle fiber bundles from EXP and measured force production *in vitro* while stimulating using three-pulse patterns with 10-10 and 12-8 ms inter-pulse intervals (IPIs). Changing the timing of the middle pulse by 2 ms significantly altered force output (Figure 3b); wfANOVA identified significant differences in the force evoked by these patterns (Extended Data Figure 4, Extended Data Table 1). Further, *d'* values evoked by 10-10 and 12-8 triplet stimulations are several standard deviations from zero (Figure 3c). The same effect holds for other pairs of similar triplets, such as 2-18 and 4-16 ms IPIs (Extended Data Figure 5a-b). Therefore, our *in vitro* experiments establish that the small, precisely-regulated differences in motor neuron spike patterns *in vivo* cause muscles to produce different forces.

We next explored whether different spike patterns not only correlate with behavior and drive distinct force production, but also *cause* different behaviors *in vivo*. We recorded air sac pressure while simultaneously applying temporally patterned electrical stimulation to EXP, again using 10-10 and 12-8 ms stimulation triplets (Figure 4a). Moving the middle pulse from 10 to 12 ms after the first evoked distinct pressure waveforms (Figure 4b) consistently across all six birds tested (Figure 4c). wfANOVA identified significant differences between the effects of these triplets (Extended Data Figure 6a, Extended Data Table 1). Finally, we comprehensively investigated the effects of moving the middle pulse from 2 to 18 ms after the first pulse (in steps of 2 ms), which resulted in significant differences in the mean pressure (Figure 4d; $p < 0.001$ for all 36 pairs of these stimulation patterns). These experiments thus demonstrate a *causal link* between millisecond-scale timing of muscle activation and the ensuing behavior.

Overall we have shown that respiratory motor unit activity is controlled on millisecond timescales, that precise timing of spikes in multispike patterns is correlated with behavior (air sac pressure), and that muscle force output and the behavior itself are causally affected by spike timing (all on similar temporal scales, Figs. 2d, 3c, 4c). These findings provide crucial evidence that precise spike-timing codes casually modulate vertebrate behavior. Additionally, they shift the focus from coding by individual spikes [3, 12, 16] to coding by multispike patterns, and from using spike timing to represent time during a behavioral sequence [17, 18] to coding its structural features. Further, we showed that the effect of moving a single spike is stable across animals (e.g., Figure 2). We believe that this precise spike timing code reflects, and exploits, muscle nonlinearities: spikes less than ~20 ms apart generate force supralinearly (Extended Data Figure 12), with stronger nonlinearities for shorter ISIs. Thus changing the first ISI from 12 ms to 10 ms not only decreases the time to the peak pressure, but also generates a larger peak (Figure 2b).

The surprising power of spike timing to predict behavior might reflect synchrony between motor units in the respiratory muscles [19], so that timing variations in one motor unit co-occur with timing variations in others. Resolving this question requires examining temporal population codes in motor systems, a subject not yet explored. Further, respiration is driven by a brainstem central pattern generator (CPG) but modified by descending inputs from the forebrain [10, 20]. It remains unknown which of these is the source of timing precision/variability. Since respiration is critical to vocalization in songbirds, it will be of especial interest to record respiratory timing patterns during singing and determine how the temporal code in upstream RA neurons [6] is transduced to the motor periphery.

Our findings suggest that current implementations of brain machine interfaces (BMIs), which typically rely on firing rates to drive action [21-23], might be improved by taking spike timing into account. Indeed, successful uses of local field potentials and electroencephalography for BMI suggest that precise (synchronized) spike timing across neurons encodes significant information about behavior [24, 25]. Extracting additional information from spike timing may thus help BMIs better decode motor activity.

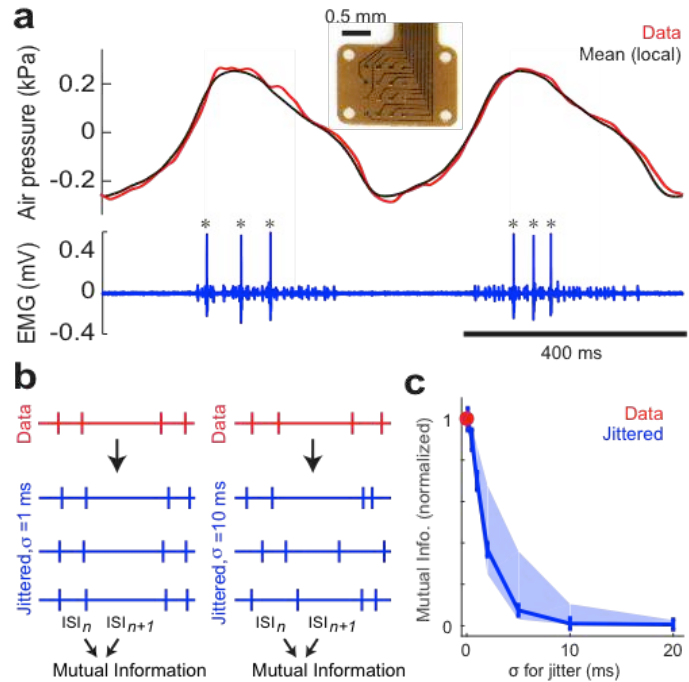

**Figure 1: The structure of the neural code.** (a) Flexible microelectrode arrays (inset) were used to record single motor units in the expiratory muscle group (EXP). A sample recording shows a well-isolated single motor unit (bottom, spikes marked by asterisks) and the corresponding instantaneous pressure and trial-averaged pressure (top, red and black respectively). (b) To identify the temporal scale of precision of spike patterns, we jittered the inter-spike intervals (ISIs) and studied the mutual information between consecutive ISIs as a function of the jitter magnitude. (c) The mutual information in jittered spike trains approached that in the original recordings only for jitters on the scale ~1 ms. Blue line shows data for Bird EMG1, for which we had the most spikes (>350,000); band shows the range across 8 anesthetized birds (EMG1-EMG8). Un-normalized values of information at $\sigma = 0$ ms were 0.057-0.146 bits.

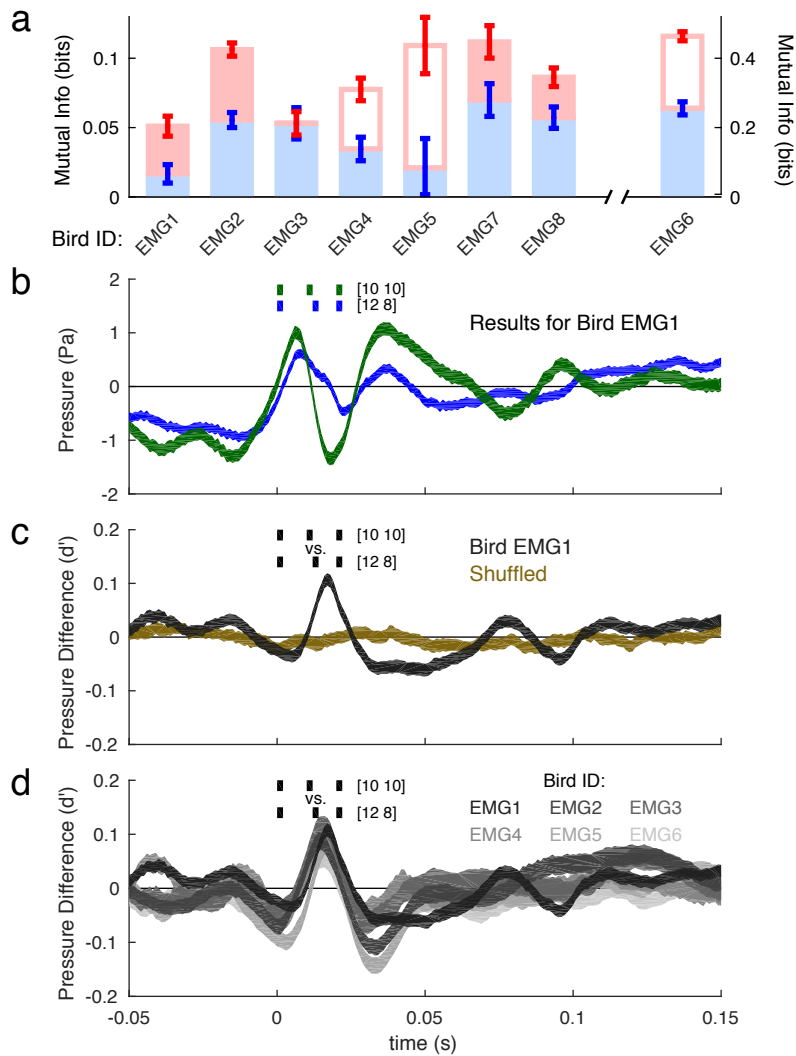

**Figure 2: Spike timing predicts respiratory air pressure.** (a) Mutual information (±1 SD) between 20 ms of spike timing (red) / spike rate (blue) and 100 ms of pressure residuals. Empty bars indicate underestimated values (see Methods). (b) Pressure residuals (the bands represent mean ±1 s.e.m., bootstrapped, see Methods) differed significantly for 10-10 (green) and 12-8 (blue) ms ISIs triplets. Negative residuals before pattern onset likely reflect activity of other correlated, non-recorded motor units. (c) Discriminability ($d'$, mean ±1 s.d., bootstrapped)

between pressure residuals for the triplets shown in (b), compared to the reshuffled control. (d) *d'* of 10-10 and 12-8 triplets in six birds (plotting conventions same as in (c)).

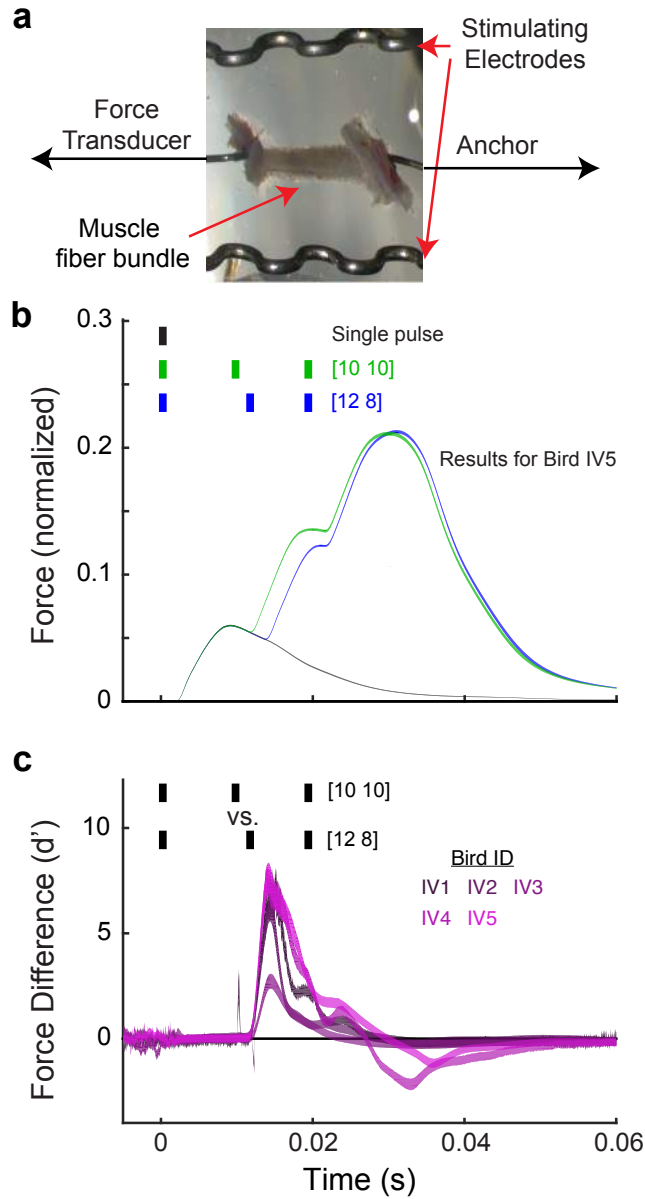

**Figure 3: Millisecond-scale stimulation patterns affect force production.** (a) *In vitro* muscle fiber preparation (see Methods). (b) Force output (mean ± s.d., bootstrapped, normalized to peak force during tetanic contraction) differs significantly between 10-10 and 12-8 ms IPI stimulation triplets (wfANOVA, Extended Data Figure 4a, Extended Data Table 1). (c) Discriminability (*d'*, mean ± s.d., bootstrapped) of force profiles in five birds (birds IV1-IV5) following stimulation with the same stimuli as in (b)

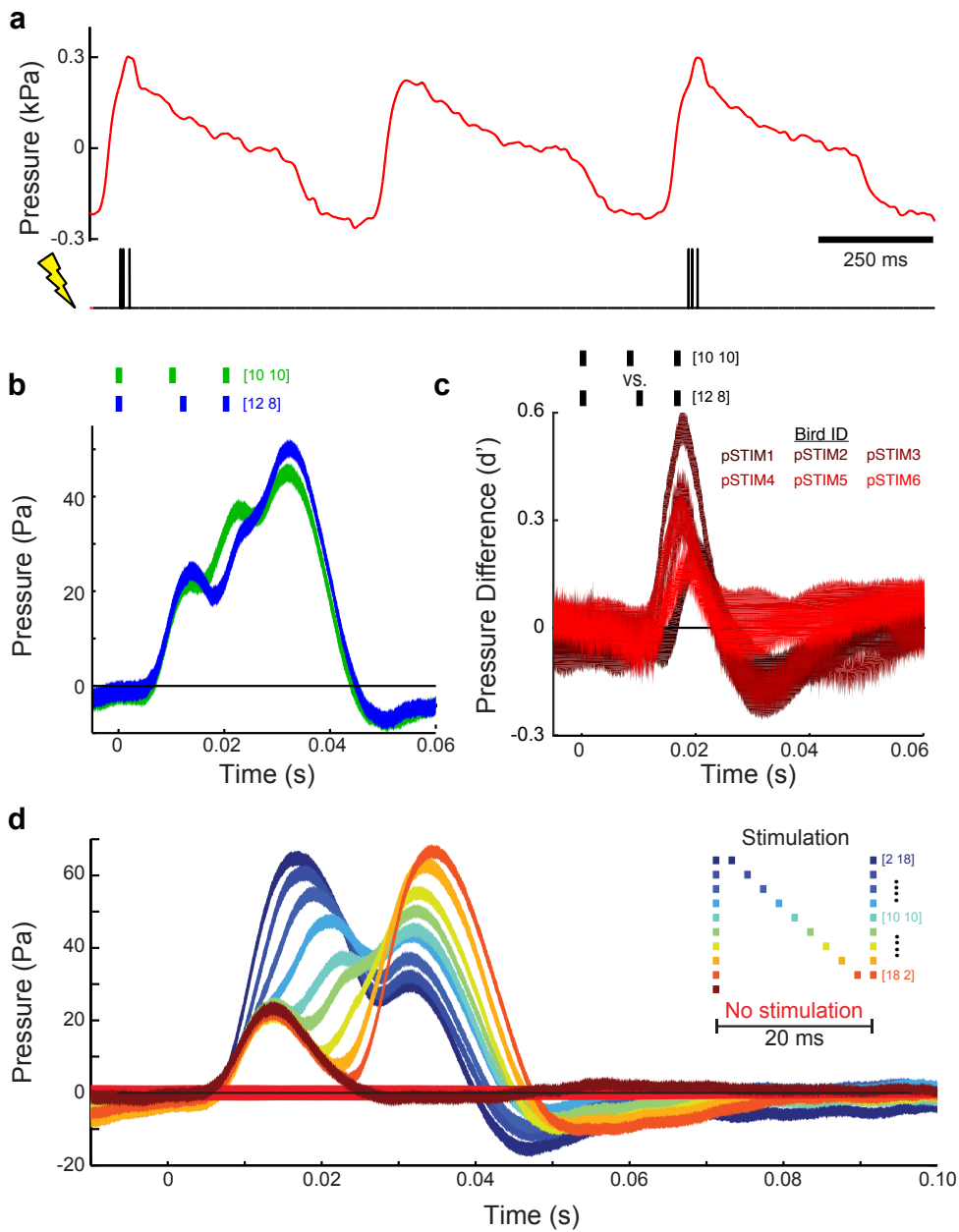

**Figure 4: Millisecond-scale differences in stimulation patterns modulate air pressure *in vivo*.** (a) Three-pulse stimulation was delivered during respiration. (b) 10-10 and 12-8 ms triplets (as in Figures 2, 3) caused significant differences in air sac pressure residuals (mean ± s.e.m., see Extended Data Figure 6a and Extended Data Table 1). (c) Discriminability (*d'*, mean ± s.d.,

bootstrapped) of pressure residuals following 10-10 and 12-8 ms triplets across birds pSTIM1-6. (d) Responses to all stimulation patterns tested (mean ± s.e.m.). In b-d, time is relative to the onset of the stimulation pattern.

1. Fairhall, A., E. Shea-Brown, and A. Barreiro, *Information theoretic approaches to understanding circuit function.* Current opinion in neurobiology, 2012. **22**(4): p. 653-659.
2. Reinagel, P. and R.C. Reid, *Temporal coding of visual information in the thalamus.* The Journal of Neuroscience, 2000. **20**(14): p. 5392-5400.
3. Rieke, F., *Spikes: exploring the neural code*. 1999: MIT press.
4. Strong, S.P., et al., *Entropy and information in neural spike trains.* Physical review letters, 1998. **80**(1): p. 197.
5. Theunissen, F. and J. Miller, *Temporal encoding in nervous systems: A rigorous definition.* Journal of Computational Neuroscience, 1995. **2**(2): p. 149-162.
6. Tang, C., et al., *Millisecond-scale motor encoding in a cortical vocal area.* PLoS Biol, 2014. **12**(12): p. e1002018.
7. Churchland, M.M., et al., *Neural population dynamics during reaching.* Nature, 2012. **487**(7405): p. 51-56.
8. Sober, S.J., M.J. Wohlgemuth, and M.S. Brainard, *Central contributions to acoustic variation in birdsong.* J Neurosci, 2008. **28**(41): p. 10370-10379.
9. Wild, J.M., F. Goller, and R.A. Suthers, *Inspiratory muscle activity during bird song.* J Neurobiol, 1998. **36**(3): p. 441-453.
10. Suthers, R.A., F. Goller, and J.M. Wild, *Somatosensory feedback modulates the respiratory motor program of crystallized birdsong.* Proc Natl Acad Sci U S A, 2002. **99**(8): p. 5680-5685.
11. Fedde, M., P. DeWet, and R. Kitchell, *Motor unit recruitment pattern and tonic activity in respiratory muscles of Gallus domesticus.* Journal of Neurophysiology, 1969. **32**(6): p. 995-1004.
12. Bialek, W., et al., *Reading a neural code.* Science, 1991. **252**(5014): p. 1854-1857.
13. Thorpe, S.J., *Spike arrival times: A highly efficient coding scheme for neural networks.* Parallel processing in neural systems, 1990: p. 91-94.
14. Paninski, L., *Estimation of entropy and mutual information.* Neural computation, 2003. **15**(6): p. 1191-1253.
15. Wickens, T.D., *Elementary signal detection theory*. 2001: Oxford University Press, USA.
16. Fairhall, A., *The receptive field is dead. Long live the receptive field?* Current opinion in neurobiology, 2014. **25**: p. ix-xii.
17. Hahnloser, R.H., A.A. Kozhevnikov, and M.S. Fee, *An ultra-sparse code underlies the generation of neural sequences in songbird.* Nature, 2002. **419**(6902): p. 65-70.
18. Leonardo, A. and M.S. Fee, *Ensemble coding of vocal control in birdsong.* J Neurosci, 2005. **25**(3): p. 652-661.
19. Chagnaud, B.P., et al., *Innovations in motoneuron synchrony drive rapid temporal modulations in vertebrate acoustic signaling.* J Neurophysiol, 2012. **107**(12): p. 3528-42.
20. Gracco, V.L. and J.H. Abbs, *Central patterning of speech movements.* Exp Brain Res, 1988. **71**(3): p. 515-526.
21. Donoghue, J.P., *Connecting cortex to machines: recent advances in brain interfaces.* Nature neuroscience, 2002. **5**: p. 1085-1088.
22. Andersen, R.A., S. Musallam, and B. Pesaran, *Selecting the signals for a brain–machine interface.* Current opinion in neurobiology, 2004. **14**(6): p. 720-726.
23. Marsh, B.T., et al., *Toward an autonomous brain machine interface: integrating sensorimotor reward modulation and reinforcement learning.* The Journal of Neuroscience, 2015. **35**(19): p. 7374-7387.

24. Waldert, S., et al., *Hand movement direction decoded from MEG and EEG.* The Journal of neuroscience, 2008. **28**(4): p. 1000-1008.
25. Ranade, G.V., K. Ganguly, and J. Carmena. *LFP beta power predicts cursor stationarity in BMI task*. in *Neural Engineering, 2009. NER'09. 4th International IEEE/EMBS Conference on*. 2009. IEEE.

**Methods:**

**Surgical Procedure**

We used electromyography (EMG) and electrical stimulation to determine the importance of motor timing in the expiratory muscle group (EXP) for avian respiration. All procedures were approved by the Emory University Institutional Animal Care and Use Committee. Prior to surgery, adult male Bengalese finches (>90 days old) were anesthetized using 40 mg/kg of ketamine and 3 mg/kg of midazolam injected intramuscularly. Proper levels of anesthesia were maintained using 0-3% isoflurane in oxygen gas.

**Subjects**

Our studies used a total of 24 adult (>90 days old) Bengalese finches. Eight birds (which we refer to as birds EMG1-EMG8) underwent EMG recordings of single motor units, in which one single unit was isolated in each animal (that is, birds EMG1-EMG8 each contributed a single unit to the study, and data from all eight are used in the analyses in main text Figure 1c). Concurrent with EMG recordings the pressure within the airsac was continuously monitored (see below for detailed descriptions of procedures). Of these eight birds, six (EMG1-EMG6) yielded sufficient neural and pressure data such that we were able to compute pressure waveforms conditional on the occurrence of a particular spike pattern (these are the birds shown in main text Figure 2a, d). Furthermore, we recorded EMG spike trains (but not air pressure, since we were unable to find a pressure meter that was both sensitive enough to detect small respiratory pressure changes and lightweight enough for an awake bird to carry) from four awake birds, which we refer to as birds EMG9-EMG12, and which provided the data shown in Extended Data Figure 1.

Additionally, muscle fiber bundles (one bundle per bird) were extracted from each of five birds for *in vitro* testing of force production. We refer to these birds (data shown in main text Figure 3) as birds IV1-IV5. We also examined the effect of electrically stimulating the expiratory muscles *in vivo* in six anaesthetized birds, which we refer to as birds pSTIM1-pSTIM6 ("pSTIM" abbreviates pressure-stimulation). Results from these subjects are shown in main text Figure 4. Lastly, four birds (C1-C4) were used to examine the effect of curare on both EMG activity and the efficacy of muscle stimulation. All birds were male except for three of the birds used in the *in vitro* experiments.

Finally, we note that three birds were used in multiple experiments. Specifically, two birds were used in both EMG unit recordings and patterned electrical stimulation, and one bird was used in both EMG recordings and (subsequently) in a curare experiment. To (possibly) reduce the complexity of our numbering scheme, the first of these subjects is referred to as both EMG7 and pSTIM1 (i.e., a single bird has two names), the second is both EMG4 and pSTIM3, and the third is both EMG8 and C1.

**EMG recordings**

To optimize our ability to isolate individual motor units, we developed micro-scale, flexible, high-density electrode arrays that sit on the surface of individual muscles to record EMG signals (Figure 1). The gold electrodes were fabricated on 20 micron thick photo-definable polyimide with a range of contact sizes and spacings (Premitec, Raleigh, NC). The electrode exposures ranged from 25 to 300 microns in diameter and were separated by as little as 25 microns. Several alignments of electrodes (16 per array) were fabricated including a four-by-four grid and four tetrodes. To record from the EXP, an incision was made dorsal to the leg attachment and rostral to the pubic bone. After spreading fascia on the muscle group, an electrode array was placed on its surface. The other end of the array connects to a custom-designed Omnetics adapter to interface with a digital amplifier (RHD2132, Intan Technologies, Los Angeles, CA). The Intan evaluation board delivered the EMG signals to the computer at 30.

With these arrays, we were able to acquire high-quality EMG recordings from 16 locations simultaneously during quiet respiration in eight male Bengalese finches. The increased number of channels allows the experimenter to decide which channels should be subtracted from each other to create bipolar signals. Because of the high specificity and impedance of individual electrodes, we were able to extract single motor unit data in some cases (Figure 1c, see Methods: Data Analysis). Because we can record sixteen unipolar signals in a very small area, we have increased the probability of recording a single motor unit while allowing us to test how different intramuscular segments are differentially recruited. Though EXP is made up of three sheet-like overlapping muscles (*m. obliquus externus abdominis*, *m. obliquus internus*, and *m. transversus abdominis*), we presume that we are recording motor units from the most superficial muscle, *m. obliquus externus abdominis*. However, since all three muscles have similar functional roles involving contraction during respiration [1], recording a motor unit from any of these muscles would not affect our interpretation.

**Pressure Recording**

Thoracic air sac pressure was monitored using a Silastic tube (Dow Corning, Midland, MI) inserted in the same manner as previous studies [2-4] with the pressure sensor 20INCH-D-4V (All Sensors, Morgan Hill, CA). Briefly, a small incision was made inferior to the rib cage. A 6 cm Silastic tube (0.03" ID, 0.065" OD) with a beveled end was inserted through the incision and sutured to the rib cage. The other end of the tube was then connected to the sensor. For recording experiments, the Intan evaluation board delivered the pressure signal to the computer at 30 kHz. For the *in vivo* muscle stimulation experiments (below), pressure recordings were acquired using a NIDAq board (BNC-2090A, National Instruments) at 32 kHz. In both cases, the Intan evaluation board provided the voltage supply for the pressure sensor.

## *In Vitro* Muscle Stimulation

The *in vitro* muscle preparation was conducted using five Bengalese finches (2 male, 3 female; none overlapping with the birds used for *in vivo* recordings and stimulation), as we have done previously [5]. We briefly describe the technique here. Animals were euthanized with an overdose of isoflurane (Baxter, Søborg, Denmark) and EXP was exposed as in the *in vivo* experiments. Fiber bundles were then isolated from the surface of *m. obliquus externus abdominis* (the most superficial muscle in the EXP group). The fiber bundles were then mounted in a test chamber while continuously being flushed with oxygenated Ringers solution at 39ºC. One end of the muscle was fixed to a servomotor (though it was not used) using silk suture while the other end was mounted on a force transducer (Model 400A, Aurora Scientific,Aurora, ON). The muscle fibers were then stimulated through the solution using parallel platinum electrodes (model 701C, Aurora Scientific). For each muscle preparation, both stimulation current and preparation length were optimized for maximum force generation. A single 300 μs pulse was used for stimulus optimization, followed by a 200 Hz, 100 ms tetanic stimulation for length optimization. To test the importance of motor timing, we stimulated the muscle with three 300 μs pulses at optimal current, with the first and the third pulse separated by 20ms, and the middle pulse being placed 2, 4, 10, or 12 ms after the first pulse. Additional trials were conducted with only a single pulse as a control. These five stimulation patterns were repeated in random order, with 60 s between each trial, and after five such iterations were followed by a 200 Hz, 100 ms tetanic stimulation. Following this procedure we measured a total of 25 iterations, taking approximately 135 min. To account for muscle fibers dying over the course of the experiment, force measurements were normalized to the fiber bundles maximum tetanic force at 200 Hz, and linearly interpolated for each stimulus. Force transducer and stimulation signals were digitized at 20 kHz with a NIDAq board (PCI-MIO-16E4, National Instruments).

*In Vivo* **Muscle Stimulation**

Stimulation of EXP was performed in six male Bengalese finches using two fine wire electrodes made of insulated multi-stranded alloy (50 μm diameter, Phoenix Wire Inc., Lawrenceville, GA). Pressure recordings were used to trigger stimulation with custom-written LabVIEW code when the pressure crossed a user-defined threshold. Generally, the stimulation pulse train was targeted for 100 ms after the pressure crossed zero. In birds where the initial upswing of the respiratory cycle was slower than normal, stimulation was delayed up to 50 ms further to prevent stimulation from occurring during the upswing itself. We wanted to avoid stimulating during that interval because it was more difficult to extract the pressure effects of stimulation. Stimulating between 100 and 150 ms after the zero-crossing also mirrored the timing of spikes found in EMG recordings. The LabVIEW code then sent a stimulation pattern to an external stimulator (Model 2100, A-M Systems, Carlsborg, WA), which was connected to the fine wire electrodes in EXP.

To test the importance of motor timing on behavior, three stimulation pulses (biphasic, 250 μs pulse duration, 250 μA current) were delivered, with the first and the third separated by 20 ms, and the middle pulse placed in 2 ms increments across the duration (9 different patterns) in addition to a single pulse and no pulse control stimuli. All 11 patterns were randomly interleaved during the experiment. Pressure and trigger times were recorded at 32 kHz using the LabVIEW code.

The selection of an appropriate current was important for interpreting the results of these experiments. To properly compare them to EMG recordings, we wanted to stimulate using a current that activates the axons of motor neurons but not muscle fibers directly. One previous study that stimulated songbird muscles used currents as great as 2 mA [5], while a more recent one posited that currents below 500 µA were likely activating nerve fibers [6]. We therefore selected a current of 250 µA for EXP stimulation for the figures shown in the main text to ensure robust effects on air sac pressure. To test the theory that we were only stimulating the axons of motor neurons, we applied curare, which locally blocks the neuromuscular junction, to EXP and compared both EMG and stimulation effects versus when saline was applied to the muscle. Curare eliminated the recorded EMG signal observed when only saline was applied to the muscle (Extended Data Figure 7a-b). Stimulation at currents as low as 100 µA produced clear effects on air sac pressure when saline was applied to the muscle (Extended Data Figure 7c), but those effects were abolished when curare was applied to the muscle (Extended Data Figure 7d). The same strong stimulation effects observed at 250 µA when saline was applied (Extended Data Figure 7e) were greatly reduced when curare was introduced to the muscle (Extended Data Figure 7f). While we were unable to completely eliminate stimulation effects at 250 µA using curare, we believe that result was due to the current spreading further than the span of the drug. Applying too much curare and fully paralyzing EXP would endanger the well-being of the animal. However, conducting our full 3-pulse stimulation experiment at 100 µA produced quite similar air sac pressure effects (Extended Data Figure 7g) as those observed using 250 µA (main text Figure 4d). Therefore, we believe our muscle stimulation experiments were only activating the axons of motor neurons and not muscle fibers directly. This allowed us to make insightful comparisons between the results of our spike pattern and stimulation analyses.

**Data Analysis**

All pressure recordings were converted from voltages to kPa via calibration measurements taken with a manometer, and then band-pass filtered between 1 and 50Hz. EMG recordings were band-pass filtered between 300 and 7,500 Hz. Pairs of EMG channels were subtracted to optimized motor unit isolation. Once a good pair was selected, motor unit spikes were sorted using custom-written MATLAB (Mathworks, Natick, MA) code [7].

To analyze the EMG and pressure recordings together (Figure 2), we searched through the pressure for occurrences of particular spiking patterns and compared the pressures following (triggered by) those patterns. We assumed that the individual spikes of our single motor unit did not drive the overall pressure cycle. So instead of comparing raw pressure measurements, we subtracted out mean cycles and analyzed statistical dependences between spiking and such pressure residuals. To define the mean waveform, we segmented individual breathing cycles (defined by a stereotypical rise and fall in pressure; segmentation by maxima, minima or zero crossings gave similar results). The cycles had different durations, and we renormalized time within each cycle to the breathing phase $(0, 2\pi)$. Since the structure of breathing changes over the hours of recordings, we averaged pressures at the same phase in a 21 cycle sliding window (other windows were tried with no significant changes), resulting in a time-dependent mean pressure waveform. The mean was then subtracted from the phase-rescaled local pressure to produce a residual in each individual cycle. We then compared these residuals triggered by the chosen spiking patterns using the usual *d'* discriminability metric,

$$d'(t) = \frac{\bar{x}(t) - \bar{y}(t)}{\sqrt{1/2\,(\sigma^2_{x(t)} + \sigma^2_{y(t)})}},$$

where $\bar{x}(t)$ and $\bar{y}(t)$ are the sample means of the pressures triggered by the two spike patterns at time $t$ after the onset of the patterns, and $\sigma_{x(t)}$ and $\sigma_{y(t)}$ are the corresponding standard deviations. To calculate the error bars on *d'*, we bootstrapped the entire analysis pipeline [8] by resampling with replacement the pressure traces residuals following the analyzed spike patterns 500 times, and estimating the standard deviation of the set of *d'* resulting from the bootstrapping. In addition to comparing patterns that are the same to two milliseconds, we also were able to compare the same patterns for a single millisecond shift in a single spike (Extended Data Figure 8), and significant differences were seen in some birds.

To isolate the effects of EXP stimulation on air sac pressure, the mean pressure waveform of the 20 previous unstimulated (catch) respiratory periods was similarly subtracted from the stimulated pressure waveform. A trailing window was used to eliminate the possibility of future effects of stimulation affecting our mean subtraction. For this pressure residual calculation, waveforms were aligned to the phase of the respiratory pattern at which the stimulation, instead of the spike pattern, occurred. All catch-subtracted pressure waveforms were averaged within a given stimulation pattern with the standard error calculated at every time point. To compare responses from two different stimuli, *d'* and its estimated error was calculated as above.

*In vitro* force measurements were compared following different stimulation patterns using the same *d'* analysis as both for the recording and the *in vitro* stimulation analyses above. Due to difficulty obtaining Bengalese finches in Denmark (where our *in vitro* studies took place), we used two male and three female subjects for analysis. Despite other experiments only being conducted on male Bengalese finches, no qualitative differences were observed between sexes, aside from normal inter-subject variability (Extended Data Figure 9). Because the sample size was small for each sex, we could not perform a statistical comparison between the two groups.

**Mutual Information: Consecutive ISIs**

In order to estimate the scale of temporal structures in the neural code, we evaluate the mutual information between subsequent ISIs that are $\leq 30$ ms long (and hence fall into the same breathing cycle), and also between these ISIs corrupted by a Gaussian noise with various standard deviations (Figure 1c). Mutual information between two continuous variables $x$ and $y$ is defined as [9]:

$$I(x,y) = \int dx\, dy\, P(x,y) \log_2 \frac{P(x,y)}{P(x)P(y)}, \tag{1}$$

and a sum replaces the integral for discrete variables. Mutual information is a measure of statistical dependency that does not assume normality of the underlying distribution in contrast to the more familiar correlation coefficient, and it measures *all* statistical dependences between the two variables, such that it is zero if and only if the two variables are completely statistically independent. Therefore, since ISIs are non-Gaussian distributed, using mutual information is more appropriate than simpler dependency measures. Mutual information is measured in bits. Measurement of *x* provides 1 bit of information about *y* (and vice versa) if the measurement of *x* allows us to answer one binary (yes/no) question about *y*.

Estimation of mutual information from empirical data is a complex problem [10, 11]. To solve this problem for mutual information between two real-valued consecutive ISIs, we use the *k*-nearest neighbors estimator [12]. The method detects structures in the underlying probability distribution by estimating distances to the *k* nearest neighbors of each data point. By varying *k*, one explores structures in the underlying data on different scales. We choose which *k* to use by calculating the mutual information for varying amount of data (using different size subsets of the full data) and detecting the (absence of) the sample size dependent bias [13-16]. The joint distribution of consecutive ISIs is smooth, and hence a broad range of $k$ near $k = 10$ produces unbiased information estimates. To identify possible sample size dependent biases and to calculate the error bars, we divided the dataset of $N$ samples into non-overlapping subsamples of size $N/m$, with the inverse data fraction $m = 2, ..., 10$. We calculated mutual information in each subsample and then evaluated the standard deviation of the estimates for a given $m$, where 10 independent partitionings where done for each $m$ (Extended Data Figure 10a). We then fitted these empirical variances to the usual 1/(sample size) law by performing a linear regression $\log \sigma^2(m) = A + \log m$. We then estimated the variance for the full data set by setting $m = 1$ (Extended Data Figure 10b). The same analysis was performed for both the original data set and the jittered data sets. To the extent that mutual information estimates for different sample sizes (different values of *m*) agree with each other within the error bars, the estimate of the mutual information likely doesn't have a sample-size dependent bias (Extended Data Figure 10a).

**Mutual Information: Spikes and Pressure**

We calculated the mutual information between 20 ms long spike trains and 100 ms long pressure residuals (Figure 2b). These timescales were chosen because spikes have to be closer to each other than about 20 ms to cause supralinear effects in muscle activation (Extended Data Figure 12) and because effects of spikes on pressure appear to last less than 100 ms (Figure 2c). We focused on spike trains that began at the phase $\varphi \approx 0.8\pi$ in the breathing cycle (see Methods: Data Analysis) because the most spiking occurs near that phase (nearby choices reveal similar results). For the pressure patterns, we started 10 ms after $0.8\pi$ phase point, since it takes about 10 ms for spiking to affect behavior (Figure 4d). Importantly, even though the starting point of patterns / pressures were chosen based on the phase, the time within the patterns and the pressure traces was not rescaled. The autocorrelation time of pressure residuals is about 10 ms for all EMG birds, so that we chose to describe the 100 ms pressure residuals using $p = 11$ real-valued data points spread 10 ms apart (see also Extended Data Figure 11a).

With this choice, we needed to calculate the mutual information between an 11-dimensional pressure vector ($y$ in Eq. (1)) and the spiking vector ($x$ in Eq. (1)). For this, we modified the $k$-nearest neighbor mutual information estimator [12] in the following way. We rewrote the mutual information between the spikes and the pressure as:

$$I(x, y) = I(n, y) + \sum_n P(n) I(x, y|n), \qquad (2)$$

where $n$ is the number of spikes in the 20 ms spike train. The first term in the r.h.s. of Eq. (2) is the information between the firing rate alone and the pressure, and the second is the information between the timing alone and the pressure. This partitioning allowed us to automatically estimate the relative contribution of each of these terms.

We estimated each of the information quantities $I(n, y)$ and $I(x, y|n)$ using the $k$-nearest neighbors estimator (recall that $I(x, y|0) = 0$), where $x$ is now an $n$-dimensional vector of spike timings for a fixed spike count $n$ (we did not discretize time for this analysis). Since mutual information is reparameterization-invariant, we rescaled the number of spikes to have zero mean and unit variance, and then additionally reparameterized the spike times and the pressure values to have normal marginal distributions, so that the $i$th value of the variable in a set of $N$ samples was mapped into the value that corresponds to the cumulative distribution of a unit variance normal being equal to $(i - 1/2)/N$. Having unit variances ensured that every variable contributes similarly to determining nearest neighbors of data points. Further, making variables to have exactly normal marginal distributions decreased the influence of outliers (we have verified that this reparameterization has a negligible effect on data with no outliers).

As in Methods: Mutual Information: Consecutive ISIs, we found the value of $k$ which produced no sample-size dependent drift in the estimate of $I(x, y)$; for EMG1, this was $k = 3$ (Extended Data Figure 11a,b). We similarly estimated error bars by subsampling the data, estimating the variance of each subsample, and extrapolating to the full data set size. For some birds, no value of $k$ produced estimates with zero sample-size dependent bias within error bars. In these cases we chose the $k$ that resulted in the smallest sample-size dependent drift. In all such cases (Bird IDs EMG3, EMG4, EMG5, and EMG6, indicated by empty boxes in Fig 2a), the drift was upwards as the sample size increased, indicating that the value of the mutual information calculated at the full sample was an *underestimate*. Underestimating the mutual information makes it harder to show that spike timing contains information about the behavior, and yet 7 out of 8 birds in main text Figure 2a showed statistically significant information in spike timing.

We additionally verified that $p = 11$ points for characterization of the pressure is the right choice by estimating $I(x, y)$ at different values of $p$. As $p$ increases from 1, more features in the pressure get sampled, uncovering more information. Information plateaus near $p \sim 10$, indicating that all relevant features in the data have been recovered, and, in some cases, it finally leaves the plateau at larger $p$ due to undersampling, indicating that such fine discretization should be avoided.

Finally, we note that $n$ in $I(n, y)$ is a discrete variable, making the use of the $k$-nearest neighbors estimator problematic. We address the issue by injecting each discrete datum with small Gaussian random noise. All data shown used the noise with standard deviation of $10^{-4}$, but other values in the range $10^{-8} \ldots 10^{-2}$ were tried with no discernable differences.

**Wavelet-based Functional ANOVA**

To compare stimulation patterns while removing inter-bird variation, we implemented an analysis technique called wavelet-based functional ANOVA (wfANOVA), which does not treat each time point in a waveform independently since comparisons are performed in the wavelet space [17]. While the above *d'* analysis provides fine temporal resolution for comparing pressure waveforms from different stimulation patterns, it treats each time point as independent from other time points, when in fact that is not the case. These two types of analyses, therefore, complement each other. The wfANOVA technique was described in detail in [17], but we outline it briefly here. Our analysis used an adapted version of the MATLAB code used in [17].

In our approach, each trial was zero-padded to make the total number of time points equal to an exponent of two, as is required for a wavelet transform. All trials were then transformed into the wavelet domain using the discrete wavelet transform, which is similar to the Fourier transform, as it produces coefficients for a family of wavelets that can be linearly transformed back into the time domain. We chose to use the third-order coiflet family, as in [17], because their symmetry does not introduce phase shifts into the data and their shapes approximated those of the pressure waveforms. The wavelets in this family are orthogonal to each other. This is important for implementing multivariate ANOVA, which performs poorly when there are high correlations between data points, as is the case in the time domain. Because wavelets are localized in time unlike Fourier components of a signal, wavelet transform of our signals featured many small or zero magnitude wavelets. The coefficients determined for each of the wavelets in the family were then used for subsequent ANOVA.

To perform the ANOVA, all trials were grouped by two factors: stimulation (or spike) pattern and bird. Fixed-effect, two-factor ANOVA was performed on the wavelet-decomposed pressure waveforms from the EMG and the stimulation experiments. This analysis therefore determined whether different spike/stimulation patterns evoked different effects on air pressure by quantifying whether one or more wavelet coefficients was significant ($\alpha = 0.05$, F test with separate post hoc Scheffé tests, with the null hypothesis being that the pressure waveforms – represented by the coefficients of the wavelet decomposition – from the pairs of stimulation patterns were equal). These post hoc tests were conducted using a significance level that was Bonferroni-corrected for the number of significant F-tests corresponding to the initial ANOVA with respect to the stimulation pattern factor (i.e., 0.05 divided by the total number of post hoc tests being conducted). Wavelet coefficients that were significantly different between stimulation patterns were subtracted to find the coefficient corresponding to the difference between the two patterns. Even having one wavelet with significant differences in their coefficients would indicate a significant difference between the two input patterns. All nonsignificant coefficients were set to zero, and the entire decomposition was transformed back into the time domain. This produced a contrast curve between two stimulation patterns across birds that removed the effects of inter-subject variability on the result. This analysis technique was also used for the *in vitro* muscle experiments, with the input being the force waveforms instead of the pressure waveforms.

# References (Methods section only)

1. Fedde, M., *Respiratory muscles.* Bird respiration, 1987. **1**: p. 3-37.
2. Wild, J.M., F. Goller, and R.A. Suthers, *Inspiratory muscle activity during bird song.* J Neurobiol, 1998. **36**(3): p. 441-453.
3. Goller, F. and O.N. Larsen, *A new mechanism of sound generation in songbirds.* Proc Natl Acad Sci U S A, 1997. **94**(26): p. 14787-14791.
4. Ashmore, R.C., J.M. Wild, and M.F. Schmidt, *Brainstem and forebrain contributions to the generation of learned motor behaviors for song.* J Neurosci, 2005. **25**(37): p. 8543-8554.
5. Elemans, C.P., et al., *Superfast vocal muscles control song production in songbirds.* PLoS One, 2008. **3**(7): p. e2581.
6. Srivastava, K.H., C.P. Elemans, and S.J. Sober, *Multifunctional and context-dependent control of vocal acoustics by individual muscles.* J Neurosci, 2015. **35**(42): p. 14183-14194.
7. Sober, S.J., M.J. Wohlgemuth, and M.S. Brainard, *Central contributions to acoustic variation in birdsong.* J Neurosci, 2008. **28**(41): p. 10370-10379.
8. Efron, B. and R.J. Tibshirani, *An introduction to the bootstrap*. 1994: CRC press.
9. Shannon, C.E. and W. Weaver, *The mathematical theory of communication*. 2015: University of Illinois press.
10. Paninski, L., *Estimation of entropy and mutual information.* Neural computation, 2003. **15**(6): p. 1191-1253.
11. Nemenman, I., F. Shafee, and W. Bialek, *Entropy and inference, revisited.* arXiv preprint physics/0108025, 2001.
12. Kraskov, A., H. Stögbauer, and P. Grassberger, *Estimating mutual information.* Physical review E, 2004. **69**(6): p. 066138.
13. Tang, C., et al., *Millisecond-scale motor encoding in a cortical vocal area.* PLoS Biol, 2014. **12**(12): p. e1002018.
14. Nemenman, I., et al., *Neural coding of natural stimuli: information at sub-millisecond resolution.* PLoS Comput Biol, 2008. **4**(3): p. e1000025.
15. Strong, S.P., et al., *Entropy and information in neural spike trains.* Physical review letters, 1998. **80**(1): p. 197.
16. Nemenman, I., W. Bialek, and R.d.R. van Steveninck, *Entropy and information in neural spike trains: Progress on the sampling problem.* Physical Review E, 2004. **69**(5): p. 056111.
17. McKay, J.L., et al., *Statistically significant contrasts between EMG waveforms revealed using wavelet-based functional ANOVA.* J Neurophysiol, 2013. **109**(2): p. 591-602.

**Extended Data / Supplementary Information**

**Analysis of regularity of spike trains**

In order to find at what timescale the spiking was controlled (and hence could be used in motor control), we looked at regularity and predictability within the single motor unit spike trains. Spikes generally occurred at a similar phase within each breathing cycle, separated by gaps on the order of hundreds of milliseconds, and so we only analyzed spikes falling within one breathing cycle. For EMG1, the mean interspike interval (ISI) for such spikes (where the same cycle was defined as ISI ≤ 30 ms) was 10.6 ms, while the standard deviation was 4.1 ms, with the refractory period of 2.5 ms. In contrast, for a refractory Poisson process with the same firing rate and refractory period, we would expect the ISI standard deviation of 7.1 ms. Thus the recorded spike train is more regular than a refractory Poisson process, suggesting that spike timing is controlled and can be used for motor coding.

**Precise timing observed in awake subjects**

In this study, we focused mostly on motor unit recordings from anesthetized, breathing birds (with simultaneous pressure recordings) in order to generate very large data sets. Several factors prevented us from performing all of our studies in awake animals. First, the pressure sensor we employed (see Methods: Pressure Recordings) was too large (10 g) to attach to an awake, freely-moving bird. While previous studies have recorded pressure chronically by attaching a smaller sensor to a backpack on a bird [2-4], we needed a sensor with better resolution in order to measure the small changes in air pressure driven by individual motor units. In contrast, previous studies often used pressure to simply identify expiration onset or qualitatively look at changes in respiration patterns. Secondly, we could not isolate single motor units over long durations of time in awake birds. We were, however, able to record single motor units in four awake birds (without concurrent pressure measurements) for shorter durations, which we used to perform an analysis similar to that done in Figure 1c in the main text. In these awake recordings, as in the anaesthetized birds, we observed similarly strong correlations between consecutive ISIs, which disappeared following jitter on the scale of 1-2 ms or larger (Extended Data Figure 1).

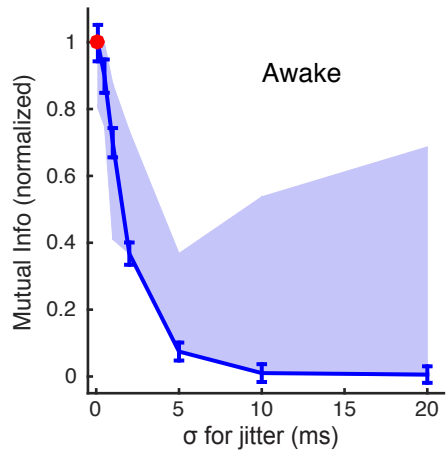

**Extended Data Figure 1: Timescales in the spike trains in awake birds.** As in anesthetized birds (main text Figure 1), mutual information in jittered spike trains recorded from awake birds approached that in the original recordings only for jitters on the scale ~1 ms. The blue line shows results from the longest spike train recorded for awake birds (EMG 9), for which we had over 9,000 recorded spikes; the band shows the range across 4 awake birds. The similarity to Figure 1c suggests that the observed millisecond-scale regularities in the spike train were not due to anesthesia. The upper limits of the range shown were driven by a bird with large error bars on its mutual information measurements (relative error of up to 27%), resulting in a statistically insignificant increase in the mutual information for larger jitters. Un-normalized values of mutual information at $\sigma = 0$ ranged from 0.094 to 0.147 bits across awake birds, which is similar to anesthetized birds.

**Individuality of motor behavior in individual birds**

Figure 2d in the main text shows that moving a single middle spike by 2 ms produces effects of a similar structure in all birds. Extended Data Figure 2 illustrates that it is only these changes that are stable across the birds, while the motor behavior itself (as quantified by PTAs) varies substantially among the birds.

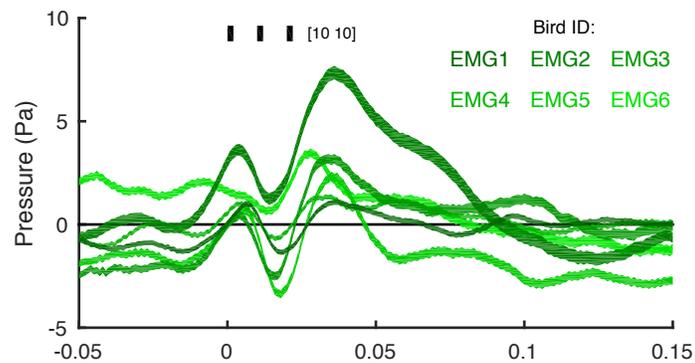

**Extended Data Figure 2: Differences in PTAs across birds.** In Figure 2d, we show that differences in PTAs across birds are remarkably consistent. This is not, however, a result of each bird having the same PTAs for each pattern, and does not mean that all of the motor units in the muscle have the same effect on air pressure. Here we show the PTAs for the 10-10 pattern treated in Figure 2; these correspond to the green trace in Figure 2b, but for all six EMG birds instead of just EMG1.

**Reconstructed wavelets from wfANOVA analysis**

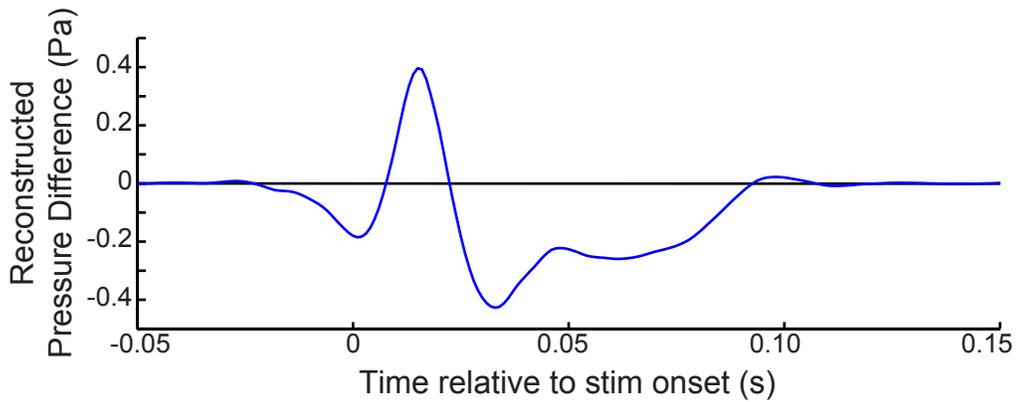

**Extended Data Figure 3: wfANOVA analysis of air pressure.** Reconstructed signal from wfANOVA (18 significant wavelets, post-hoc $\alpha = 6.66 \times 10^{-4}$) revealed the difference in air sac pressure between 3-spike patterns over 20 ms, with the middle spike either occurring at 10 ms or 12 ms after the first spike (i.e., 10-10 vs 12-8 spike triplets), independent of inter-subject variability (see main text Figure 2b-d).

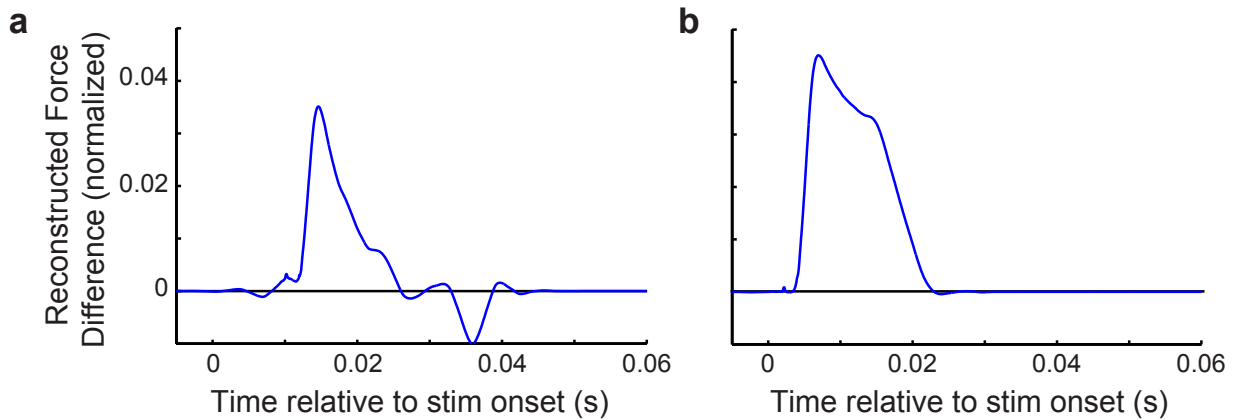

**Extended Data Figure 4: wfANOVA analysis of *in vitro* muscle force.** Reconstructed signal from wfANOVA revealed the effect of moving the *in vitro* stimulation pulse on muscle force, independent of inter-subject variability. Note that as in main text Figure 3b, force measurements are normalized to the maximum force during tetanic contraction. For 3-pulse *in vitro* stimulations taking place over 20 ms (see Figure 3), significant wavelets demonstrated the difference between patterns with the middle pulse at (a) 10 ms and 12 ms (i.e., 10-10 vs. 12-8 triplets, 29 significant wavelets, post-hoc $\alpha = 2.71 \times 10^{-4}$) and at (b) 2 ms and 4 ms (2-18 vs. 4-16 stimulation triplets, 47 significant wavelets, post-hoc $\alpha = 2.97 \times 10^{-4}$).

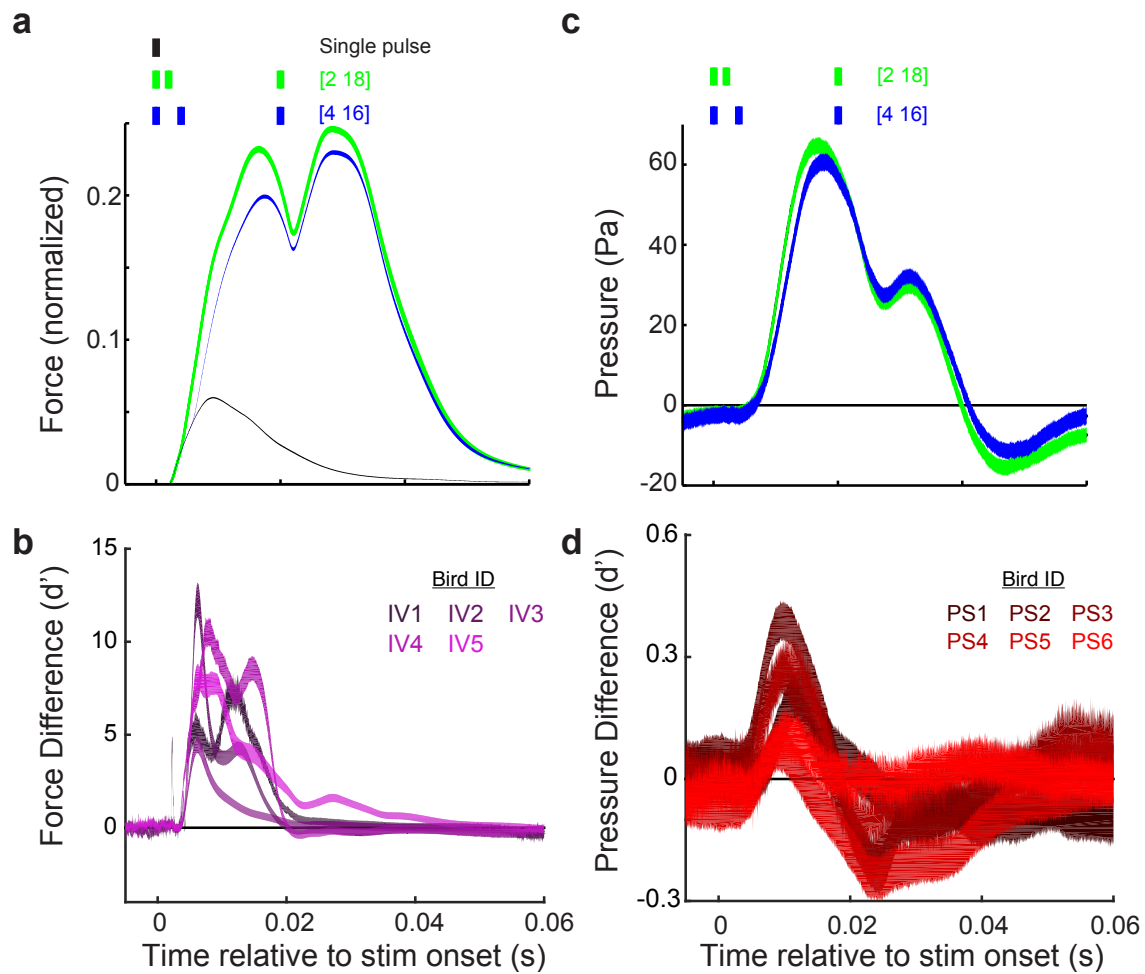

**Extended Data Figure 5:** Motor timing showed significant effects on force and behavior when middle pulse is close to the beginning of interval (in the main text, we studied cases when the middle pulse was near the middle of the interval). (a) Force waveforms were significantly different when moving the middle stimulation pulse from 2 ms to 4 ms after the first spike in a 20 ms 3-spike pattern (47 significant wavelets, post-hoc $\alpha = 2.97 \times 10^{-4}$, see Extended Data

Figure 4b). (b) This significant effect, as measured by $d'$, was observed across all 5 subjects. (c) The same stimulation patterns caused a significant difference in air sac pressure (16 significant wavelets, post-hoc $\alpha = 1.64 \times 10^{-4}$, see Extended Data Figure 6b below). (d) This effect, measured by $d'$, was also observed across all 6 subjects. Note that these stimulation patterns tested the limits of the system, but they were not observed as spike patterns during EMG recordings in unstimulated birds.

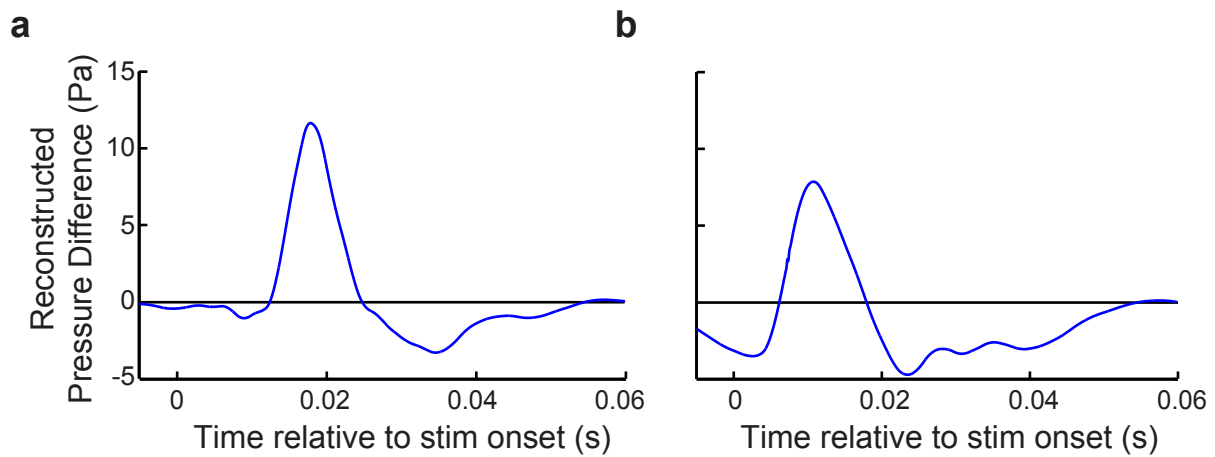

**Extended Data Figure 6:** Reconstructed signal from wfANOVA revealed the effect of moving the *in vivo* stimulation pulse on air sac pressure, independent of inter-subject variability. For 3-pulse *in vivo* stimulations taking place over 20 ms (main text Figure 4), significant wavelets demonstrated the difference between patterns with the middle pulse at (a) 10 ms and 12 ms (i.e., 10-10 vs. 12-8 pulse triplets, 14 significant wavelets, post-hoc $\alpha = 1.18 \times 10^{-4}$), and at (b) 2 ms and 4 ms (i.e., 2-18 vs. 4-16 triplets, 16 significant wavelets, post-hoc $\alpha = 1.64 \times 10^{-4}$).

**Muscle stimulation activates motor neurons**

Another important consideration for this study is how electrical stimulation actually excites the muscle tissue. Electrical stimulation of muscle tissue (see Methods) might evoke muscle contraction in one of two (non-exclusive) ways: either by directly depolarizing muscle fibers and/or by inducing action potentials in the motor neurons innervating the muscles. To distinguish these possibilities, we examined the effect of applying curare (which blocks synaptic transmission at the neuromuscular junction). As expected, curare treatment largely abolished EMG signals (Extended Data Figure 7a,b). We then compared the effects of patterned electrical stimulation on air pressure with and without curare. We found that curare completely abolished the air pressure changes induced by 100 μA current (Extended Data Figure 7c,d) and mostly abolished air pressure changes induced by 250 μA current (Extended Data Figure 7e,f). Therefore, although we cannot rule out some contribution of direct muscle fiber activation in our *in vivo* stimulation experiments, it seems likely that any such contribution is minimal. Nevertheless, to evaluate whether such a contribution of direct muscle fiber activation might account for the results shown in main text Figure 4d, we repeated this experiment with 100 μA stimulation current. As shown in Extended Data Figure 7g, this smaller current, the effect of which is entirely mediated by motor axon activation, produces qualitatively identical results as the 250 μA current used in Figure 4d, providing strong evidence that our *in vivo* stimulation paradigm affect behavior via neuromuscular transmission.

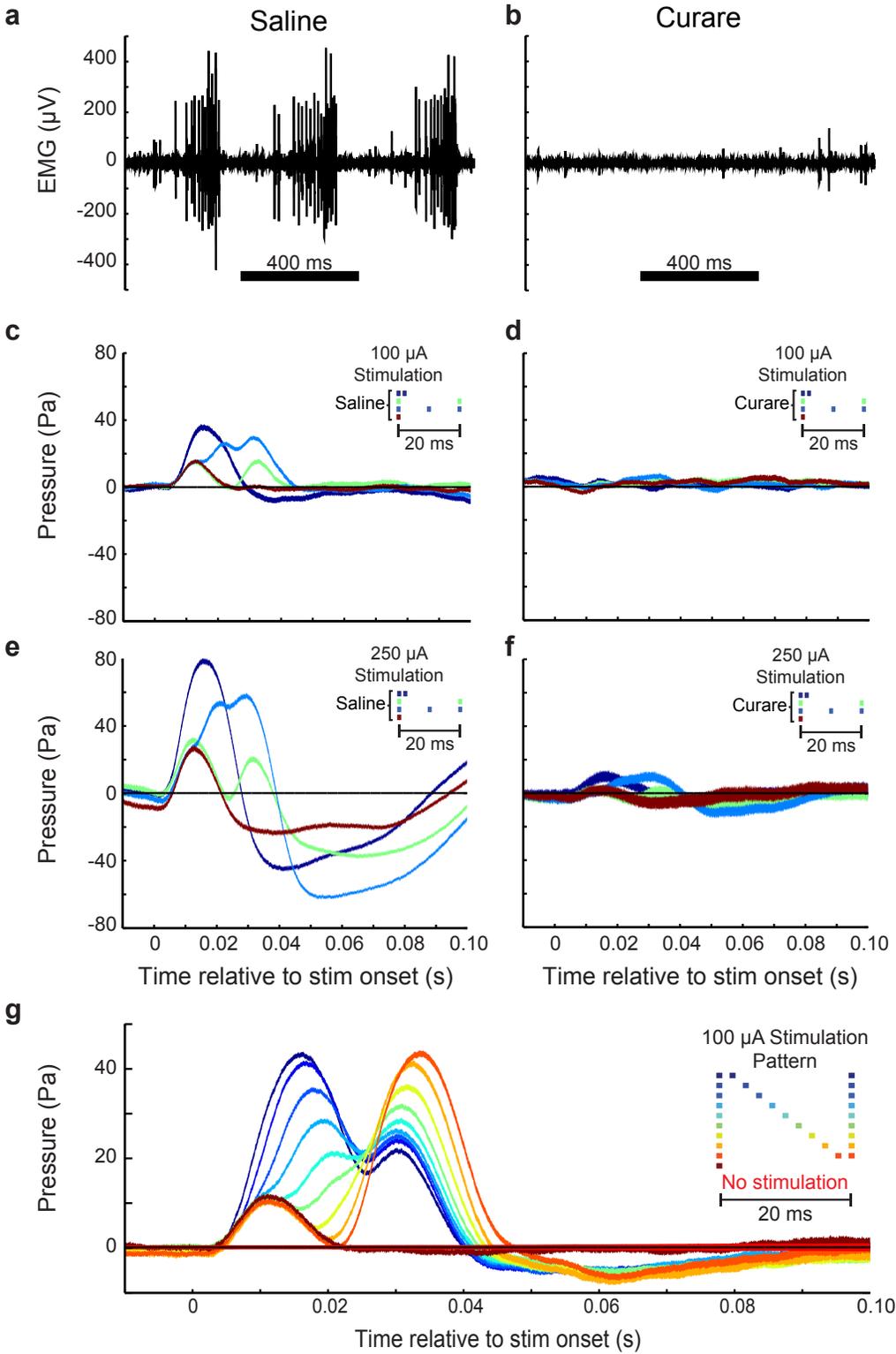

**Extended Data Figure 7: Curare experiments suggest that EXP stimulation activates the axons of motor neurons, and not the muscle fibers directly.** (a) EMG recordings showed

strong activity when saline was applied to the muscle, (b) but that activity quickly disappeared when curare, which locally blocked synaptic transmission at the neuromuscular junction, was applied to the muscle. (c) Various stimulation patterns at 100 μA had clear effects on air sac pressure, (d) but curare abolished those effects at the same stimulation current. (e) The same stimulation effects at 250 μA were (f) greatly reduced by curare. (g) Though our experiments in the main text described effects of EXP stimulation at a current of 250 μA, we saw similar effects at currents, where stimulation effects were abolished by curare as in (c), such as 100 μA shown here. The color bars represent mean +/- s. e. m.

**Even triplets differing by 1 ms can predict different behaviors**

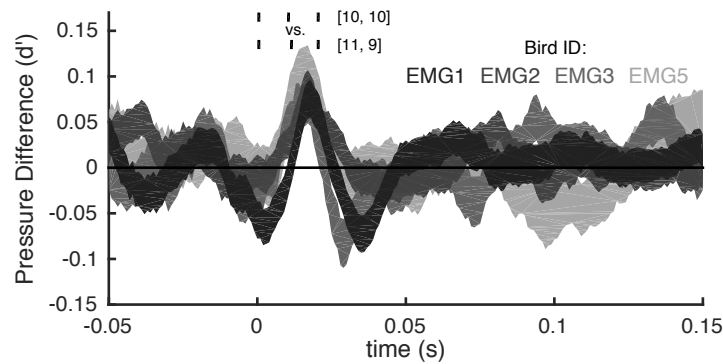

**Extended Data Figure 8: Distinguishability of Pattern Triggered Averages (PTAs) on 1 ms scale.** Similar to main text Figure 2d, here we plot *d'* (+/- s.e.m, bootsrapped) for the distinguishability of the pressure traces triggered by 10-10 and 11-9 ms ISI patterns — different only by the position of the middle spike in a spike triplet by 1ms. Of the six birds analyzed in Figure 2d, four (EMG1, EMG2, EMG3, and EMG5) had statistically significantly nonzero *d'* curves triggered by these 1-ms different spike patterns. At peak, EMG1 showed *d'*=0.08+/-0.02. The decrease in statistical significance likely reflects having many fewer equivalent patterns if viewed at a higher temporal resolution ($N$ = 23,991 and 6,507 for 10-10 triplet for EMG1 at 2 ms and 1 ms resolution, respectively).

## *In vitro* stimulation results do not vary by sex

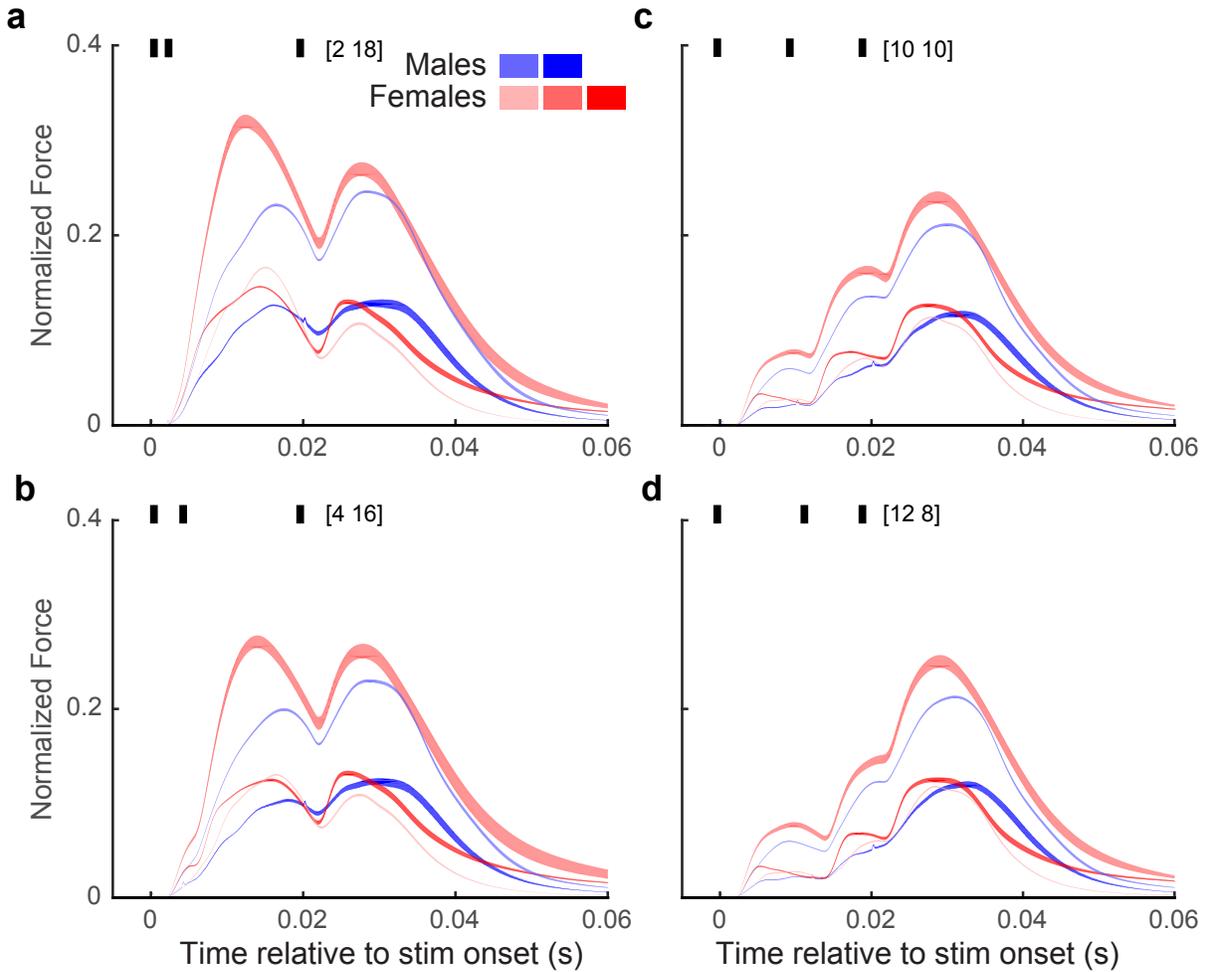

**Extended Data Figure 9:** *In vitro* **stimulation of EXP did not show any qualitative differences across sexes.** Across all four stimulation patterns (a-d), all subjects showed similar force trajectories, with no subjects showing differences outside of normal inter-subject variability. Sample size for each group was too small to perform a statistical comparison, but the data shown above suggest that using muscle fibers from both male and female birds for *in vitro* studies (see *Methods*) did not greatly affect our results.

**Mutual Information Estimation**

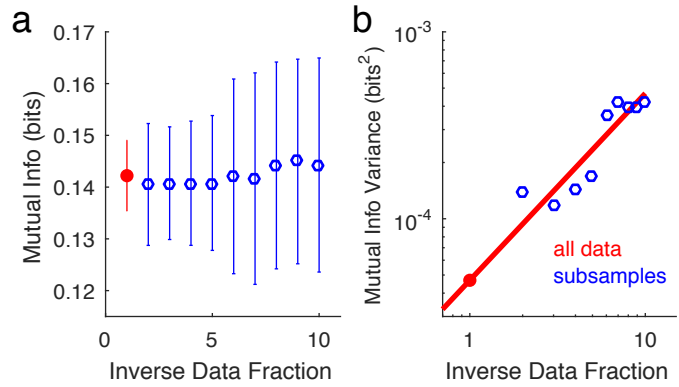

**Extended Data Figure 10: Estimation of mutual information between consecutive ISIs for EMG1.** (a) The plot of mutual information vs. data set size shows stability of the calculated values, and hence absence of the sample-size dependent bias in the estimation of information. The full data set here includes $N = 154{,}548$ consecutive ISIs pairs (red dot). Estimates for other data set sizes (blue circles) are obtained by taking $1/m$ fraction of the total amount of data, with $m$ shown on the horizontal axis (see *Methods: Mutual Information: Consecutive ISIs*). (b) For subsampled data (blue), we estimate the variance, $\sigma^2$, of the mutual information estimates ($\sigma$ is also shown as error bars in panel (a)) as the variance of $m$ non-overlapping subsamples. We repeated this ten times at each $m$ and averaged the variances. As discussed in Methods: Mutual Information: Consecutive ISIs, we then extrapolated the variance to the full data set using linear regression (red line and dot), and the extrapolated standard deviation is shown as the error bar in panel (a).

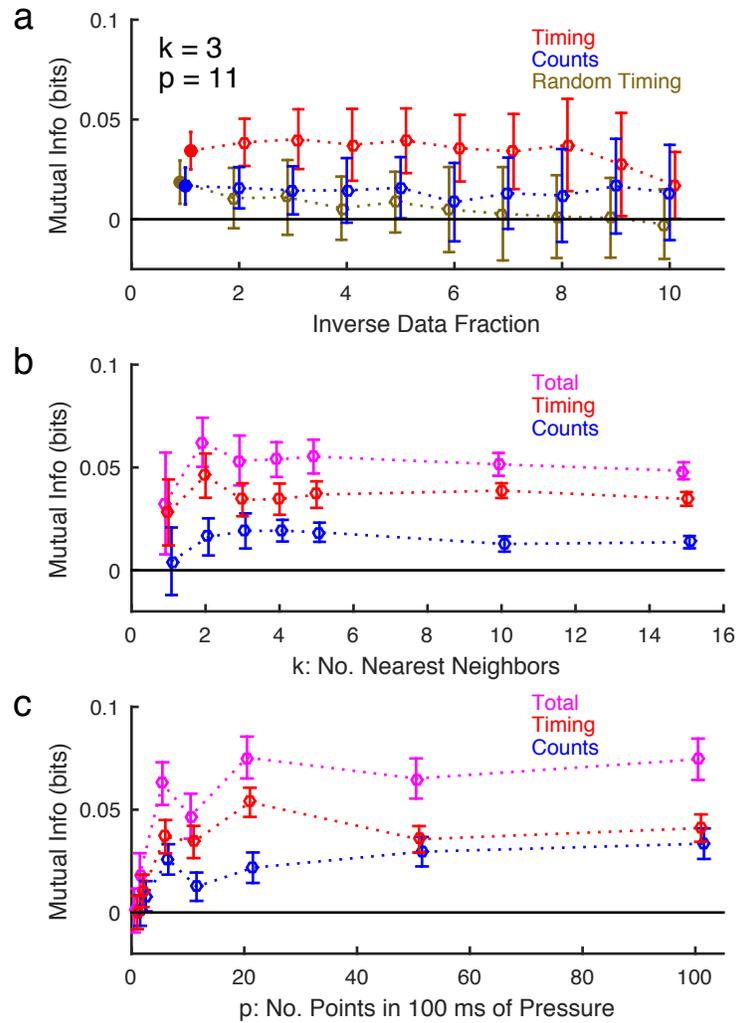

**Extended Data Figure 11: Estimation of mutual information between spikes and pressure for EMG1.** (a) We plot mutual information as a function of inverse data size, similar to Extended Data Figure 10a for the chosen $k$, the number of nearest neighbors in the mutual information estimator (see panel (b)), and $p$, the number of data points used to represent the pressure (see panel (c)). Here the full data set is $N$=16,516 breathing cycles (solid circle, inverse data fraction of 1). Information between the spike count and the pressure (blue) and the spike timing and the pressure (red) are shown separately (also reproduced in main text Figure 2a for $p = 11$, $k = 3$, and the full data set). To further establish statistical significance of the timing mutual information estimate, we show the mutual information between the pressure traces and randomly

shuffled spike timings, while keeping the number of spikes fixed (brown). This information falls within error bars from zero, as it should because random spike timing does not carry information about the pressure. (b) As explained in Methods, we chose the number of nearest neighbors, $k$, for the estimation algorithm ($k = 3$ for EMG1) such that $k$ is large enough to have small estimated error bars, and yet small enough, so that fine features in the distribution are not averaged out, and the information does not decrease a lot from the maximum. We verify the choice of $k$ by insisting on the smallest possible sample size dependent drift in the estimated information, as in panel (a). Spike timing (red), spike count (blue), and total (magenta) mutual informations are shown. (c) As elaborated on in Methods, we chose the same number of points ($p = 11$, 10 ms apart) to represent 100 ms of pressure in all EMG birds. The choice was precipitated by using the smallest $p$ (which leads to best sampling and smallest overall error bars), such that the information reached its large-$p$ plateau across all birds within error bars. As in panel (b), spike timing (red), spike count (blue), and total (magenta) mutual informations are shown.

**Determining relevant time intervals**

To determine the salient duration to test 3-pulse stimulation experiments, two stimulation pulses (each pulse was biphasic with a 250 μs pulse width at 250 μA) were delivered with the interpulse interval (IPI) varied to include 1, 2, 3, 4, 5, 6, 7, 8, 9, 10, 12, 14, 16, 18, 20, 25, 30, 40, and 60ms. As controls, a single pulse and a null stimulus were also delivered. All 21 patterns were interleaved during the experiment. We suspected that a precise timing code would rely on muscle nonlinearity at short interspike or interpulse intervals to produce significant differences in forces due to changes in spike timing on the scale of a few milliseconds. Thus we designed experiments to find interpulse intervals that would have a nontrivial, supralinear effect on the motor output. Such interpulse and interspike intervals (≤ 20 ms in length, Extended Figure 12) were then used in the rest of the research (e.g., 10-10 and 12-8 ISI patterns in main text Figure 2).

To find the characteristic time scale for supralinear force production, we measured nonlinear summation by comparing the actual pressure response to a given stimulation pattern with the response constructed by adding the pressure response to a single stimulation pulse at the same times. At a small IPI of 2 ms, the nonlinear summation was large (Extended Data Figure 12b), while the difference was negligible at an IPI of 20 ms (Extended Data Figure 12c). We averaged the area difference between these two responses across trials, then took the absolute value for IPIs between 1 ms and 60 ms. This metric showed that the supralinear effect disappeared at IPIs above 20ms (Extended Data Figure 12d).

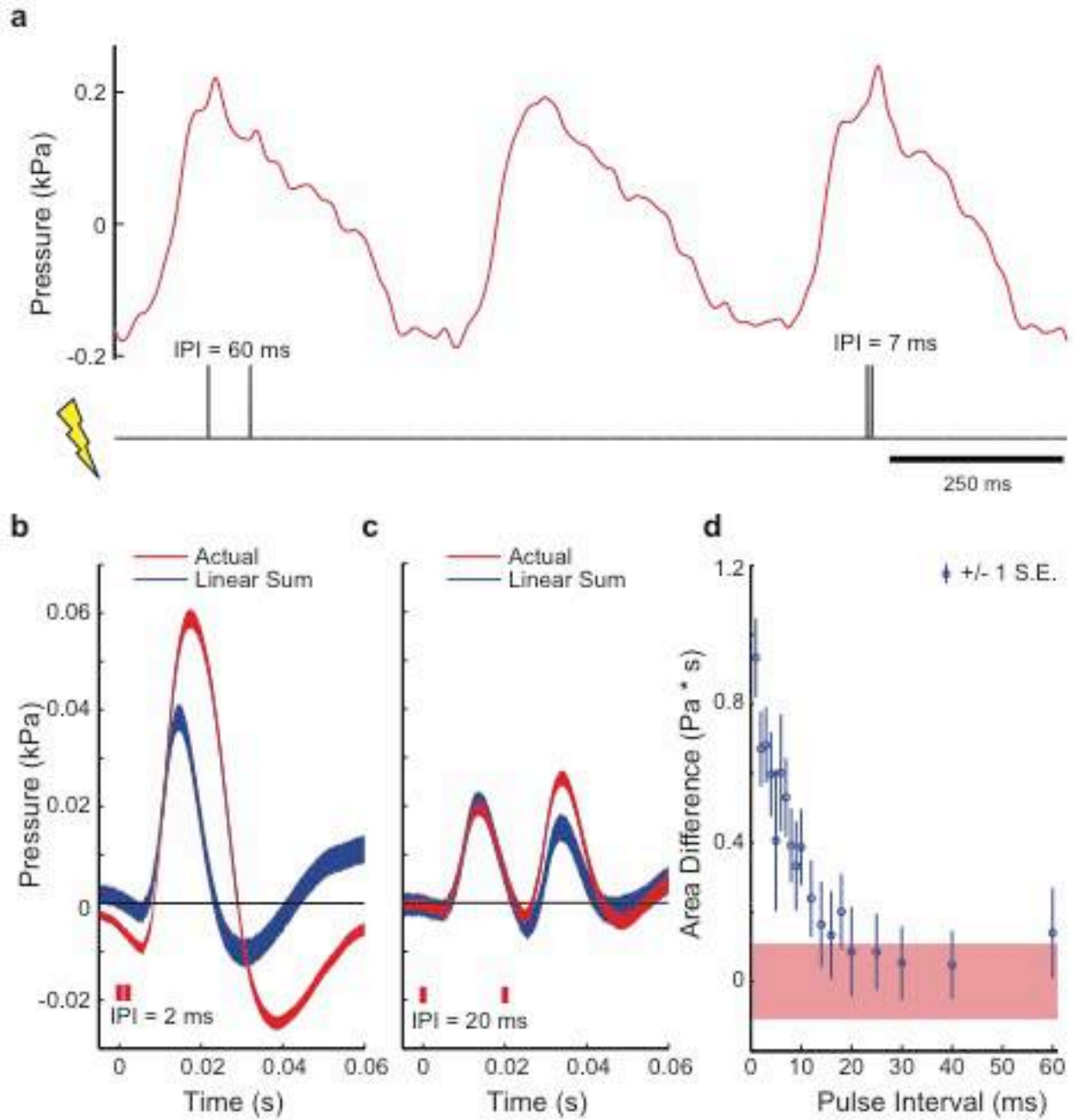

**Extended Data Figure 12: Nonlinear summation in thoracic air sac pressure in response to *in vivo* stimulation.** (a) During respiration, stimulation of a varying interpulse interval (IPI) was delivered after air sac pressure crossed a user-defined threshold (see *Methods: In Vivo Muscle Stimulation*). (b) At small IPIs ~2 ms, the actual pressure response was much greater than that produced from summing two single pulse responses separated by 2 ms. (c) On the other hand, the responses were quite similar for an IPI of 20 ms. In these plots, the width of the color bands represent the mean +/- s.e.m. (d) The difference between the actual and the summed response dropped as IPI increased, leveling off after 20 ms. The red band denotes the mean +/- s.e.m. when calculating the area difference between the single pulse responses and the mean single pulse response; we expect the area differences between equivalent stimulations to be on this scale as well. Note that the supralinear summation was relatively small for a 1 ms IPI, as the second pulse likely occurred within the motor neuron's refractory period. Data shown are from bird pSTIM3.

**Extended Data Table 1: Results of wfANOVA**

| Type of Experiment | Number of Significant Wavelets | Bonferroni-Corrected Alpha | Smallest p-value | Largest p-value | Figure Reference |
|---|---|---|---|---|---|
| EMG | 18 | $1.10 \times 10^{-3}$ | $3.36 \times 10^{-86}$ | $6.66 \times 10^{-4}$ | Fig. 2, Extended Data Fig. 3 |
| *In vitro* stim (10-10 vs 12-8) | 39 | $3.70 \times 10^{-4}$ | $4.04 \times 10^{-105}$ | $2.71 \times 10^{-4}$ | Fig. 3, Extended Data Fig. 4a |
| *In vitro* stim (2-18 vs 4-16) | 47 | $3.47 \times 10^{-4}$ | $3.47 \times 10^{-82}$ | $2.97 \times 10^{-4}$ | Extended Data Figs. 4b, 5a-b |
| Air Sac Stim (10-10 vs 12-8) | 14 | $1.18 \times 10^{-4}$ | $5.31 \times 10^{-56}$ | $6.79 \times 10^{-5}$ | Fig. 4, Extended Data Fig. 6a |
| Air Sac Stim (2-18 vs 4-16) | 16 | $1.64 \times 10^{-4}$ | $2.73 \times 10^{-50}$ | $1.43 \times 10^{-4}$ | Extended Data Figs. 5c-d, 6b |

**Conflict of interest:** The authors declare no competing financial interests.

**Acknowledgements:** This work was supported by National Institutes of Health grants P30NS069250, R01NS084844, F31DC013753, and 5R90DA033462, National Science Foundation grant 1208126, James S. McDonnell Foundation grant 220020321, Danish Research Council (FNU) and Carlsberg Foundation, and the Woodruff Scholarship at Emory University. We thank Lucas McKay for his help using wavelet-based functional ANOVA.

**Author contributions:** KHS designed research, performed experiments, analyzed data, wrote paper; CMH analyzed data, wrote paper; MV and AP performed experiments; CPHE designed research, performed experiments, wrote paper; IN and SJS designed research, analyzed data, wrote paper.